# High-order Discontinuity Detection Physics-Informed Neural Network

Ruquan You [1,2] & Shuming Zhang [1,2] & Tinglin Kong[1,2], Haiwang Li [1,2,*]

[1)] Research Institute of Aero-Engine, Beihang University Beijing, 100191, China

[2)] National Key Laboratory of Science and Technology on Aero Engines Aero-thermodynamics, Beihang University Beijing, 100191, China

(*Corresponding Author: lihaiwang@buaa.edu.cn)

## Abstract

In order to solve the problem of the difficult direct measurement of temperature field in fluid machinery under high-speed compressible conditions, this study combines high-order finite difference numerical format, Weighted Essentially Non-Oscillatory (WENO) discontinuity detection, and traditional Physics-Informed Neural Network (PINN) to develop a high-order discontinuity detection PINN (Hodd-PINN) that can achieve temperature field inversion with a small number of measurement points. When dealing with pure convection problems, Hodd-PINN introduces a 7th-order discretization for the convection term, reducing an additional 9.7% error compared to traditional low-order discretization methods. When dealing with pure diffusion problems, Hodd-PINN introduces an 8th-order discretization for the diffusion term, reducing an additional 12.8% error compared to traditional low-order discretization methods. In addition, this paper develops a loss function based on WENO discontinuity detection technology, which helps eliminate false discontinuities, allowing Hodd-PINN to successfully identify sparse waves that are easily overlooked in PINN's predicted results, reducing the error by 24.2%. Through extensive testing, this paper points out that the Hodd-PINN, which incorporates high-order discretization and discontinuity detection technology, can further reduce the prediction error of PINN, effectively reducing the data requirement, and can effectively solve the problem of false discontinuities. This method has important value for the inversion of temperature and velocity fields in fluid machinery under high-speed compressible conditions.

## Introduction

In the field of high temperature, intense aerodynamic compression, and strong heat transfer[1][2][3][4], temperature control has become the main challenging aspect in scientific research and industrial design. However, the high temperature, high speed, and strong heat transfer characteristics make it extremely difficult to conduct full temperature and pressure experiments [4][5], resulting in typical problems such as limited temperature testing methods. Currently, temperature detection methods are mainly divided into contact measurement and non-contact measurement. Among them, contact measurement has limitations due to the influence of high temperature and high-speed airflow, requiring probes to have good heat protection in order to function properly [6][7]. Moreover, considering the interference from complex flow, obtaining valid experimental results requires extra effort [8]. Non-contact measurement mainly includes infrared temperature measurement [9], liquid crystal temperature measurement [10], and thermocouple temperature measurement [11]. Among them, infrared temperature measurement can obtain the surface distribution of temperature, but due to the non-uniform characteristic of radiation with angle, it has a larger measurement error. Liquid crystals are sensitive to temperature, but their physical properties make it difficult to withstand full temperature and pressure experiments, and the sprayed thickness can also have an unpredictable impact on the heat transfer characteristics of the test piece. In most cases, if precise measurement of the surface temperature of the experimental component in a high temperature and high-speed environment is desired, it is usually necessary to use embedded thermocouples. Although the process of embedding thermocouples is complex and the spatial resolution is extremely limited (typically only a few measurement points can be set up), accurate measurement results can be obtained. However, the measurement method of embedded thermocouples is difficult to obtain the complete temperature field of the experimental component, which is very unfavorable for fine temperature control design. Therefore, it has always been an important issue to use as few temperature measurement points as possible to obtain the full temperature field of the entire experimental component, and to achieve temperature field inversion based on limited experimental data.

Traditional CFD methods are difficult to integrate experimental data. On one hand, traditional grid-based CFD methods apply control equations to corresponding computational domains through discrete solution methods [12], and use Taylor expansion to control the error limit between discrete points and continuous physical fields, and then determine the solution based on known boundary conditions [13][14][15][16]. On the other hand, once the boundary conditions are determined, the discrete solution has already reached global convergence and cannot accept secondary modifications. From another perspective, "having a determined solution" and "accepting modified computational results" are two contradictory properties that cannot be simultaneously satisfied. Therefore, although traditional CFD methods can solve physical fields based on boundary conditions, they cannot further integrate the data. However, methods like PINN [17] and TGNN [18], which combine CFD and neural networks, provide new possibilities for integrating experimental data. In a narrow sense, PINN is a method that uses neural networks to fit continuous physical fields, avoiding the process of obtaining physical fields through discrete solutions [19]. In the coordinate positions where the measured values of the physical field are known, PINN can directly construct a function from the local coordinates to the predicted values. In the coordinate positions where the measured values of the physical field are unknown, PINN can rely on the constraints of partial differential equations for training without data, so that the trained positions can satisfy the partial differential equations to the greatest extent possible, achieving physical field predictions within a certain range of accuracy [20][21][22]. However, since the spatial positions used for training PINN are always limited, the NN always needs to generalize the properties of these trained points to the surrounding space. This approach means that PINN can accept both measurement data or differential equation laws at observed positions, and can flexibly choose the direction of generalization at unobserved positions. Because of PINN's reliance on generalization for solving, strictly speaking, PINN does not have a determined solution, only a set of solutions that satisfy the given observation point rules. From this perspective, the reason why PINN can achieve data fusion is that it only needs to optimize and find a solution set that is closest to the given constraints, without guaranteeing that its own prediction results hold true at every position in space. Due to the existence of the universal approximation theorem [23], as long as the solution has continuous characteristics and the expressive power of the NN is strong enough, such conditions can always be satisfied. Therefore, PINN achieves the flexibility of data fusion, which can be seen as sacrificing determinism in order to gain benefits.

Due to the inherent nature of PINN mentioned above, PINN in a narrow sense has deficiencies. These deficiencies arise from the way physical constraints are added and the general instability of neural networks. On one hand, although PINN incorporates physical constraints into the loss function, the loss function itself may not be fully satisfied, or it may easily get trapped in local optima, resulting in PINN's predicted results not being completely consistent with the solutions of partial differential equations [24]. On the other hand, the generalization ability of PINN serves as a bridge between discrete data and continuous physical fields. Its generalization is established statistically and cannot guarantee that the position between two training points will not fail [25]. Such instability makes PINN's predicted results less rigorous. In fact, if PINN uses Relu[26] or LeakyRelu[27] as the activation function, the derivative mutation caused by generalization instability will become exceptionally noticeable. In order to compensate for the deficiencies in generalization ability as much as possible, PINN needs to rely on a large number of data points to fit the physical field, similar to traditional neural networks [28]. Even so, without a large number of experimental measurements for correction along the way, training only relying on partial differential equation constraints may still result in the accumulation of errors in space, making extrapolation performance difficult to guarantee. Apart from the high requirements for training data and methods, traditional PINN also has theoretical deficiencies when faced with physical problems with discontinuities: PINN assumes that in space points with unknown exact values, the physical field should satisfy the partial differential equations. However, when facing nonlinear problems or even discontinuous physical fields with discontinuities caused by compressible flow, such as shock waves, the assumption of being continuously differentiable will locally fail, leading to a deterioration in the robustness of PINN [29].

To solve the inherent shortcomings of Physics-Informed Neural Networks (PINN), scholars have made improvements from two main perspectives. The first perspective is to expand the traditional PINN methods from the point of view of statistical learning. For example, Qiu et al. [30] utilized transfer learning to first learn a temperature field under similar operating conditions, and then supplemented the training with a small amount of measurement data, thereby transferring the original fitting results to the new temperature field. This method partially solves the problem of insufficient effective measurement data in real turbine blade experiments. Zhang D et al. [31] designed multiple deep neural networks (DNNs) to learn orthogonal bases of arbitrary random spaces, introducing uncertainties in the parameters and approximations of DNN, which improves the predictive performance of DNN. Jagtap A D et al. [32] proposed a method to adaptively adjust the activation function, which can approximate smooth and nonsmooth functions as well as linear and nonlinear partial differential equation solutions, significantly improving the convergence speed. Zheng J et al. [33] proposed a spatiotemporal hybrid model (STHN) that combines Convolutional Neural Networks (CNN) [34] and Bidirectional Long Short-Term Memory Networks (BLSTM) [35][36]. Compared with traditional empirical and regression models, this method has better predictive performance and higher accuracy. These methods optimize the target problems in the PINN field from the nature of neural networks themselves, but due to the black-box nature of neural networks, these methods often focus on specific neural network designs and optimize the problems less from a physical principle perspective.

In addition to expanding the traditional PINN method from the perspective of statistical learning, another approach is to integrate discrete numerical formats and similar ideas with traditional PINN, thereby providing further expert constraints. For example, Xu K et al. [37] proposed a method that combines neural networks and numerical methods for solving partial differential equations, achieving faster convergence speed and better stability. This method can be used to solve a wide range of data-driven inversion models and obtain accurate solutions. Xu R et al. [38] introduced a Weak Form Theory-guided Neural Network, which transfers the higher-order derivatives in the PDE by partially integrating them into the test function. With the help of the integration concept, they implemented a low-pass filter at the level of the loss function, making it more robust to discontinuities and noise. Mohan A T et al. [39] proposed a method that enforces the introduction of physical constraints and boundary conditions using a convolutional neural network architecture with strong inductive bias, thereby embedding the principle of incompressibility for fluid dynamics. Xiang Z et al. [40] presented a method called HFD-PINN, which is a hybrid of physics-informed neural network and mixed finite difference method. It not only achieves faster convergence, but also provides better prediction results. Chen Y et al. [24] proposed a theory-guided hard constraint projection method (HCP), which introduces hard constraints by projecting the PINN computation results into the expression space of the discrete numerical format. The uniqueness of the solution is mathematically proven. The inherent problems of traditional PINN include high data demand and unreliable generalization, which can be remedied by introducing traditional finite volume and finite difference methods. This can be interpreted as further constraining the predictive error using Taylor expansions. Traditional PINN relies solely on generalization to bridge the gap between continuity and discreteness. By adding constraints based on Taylor expansions, it also introduces strict relationships between multiple discrete points. Therefore, this type of PINN not only possesses the advantages of traditional numerical format with finite error, but also retains the data fusion capability of PINN.

Although the methods mentioned above, which introduce physical constraints, effectively improve the performance of Physics-Informed Neural Networks (PINN), they still face common challenges when dealing with more complex problems such as transonic flows, sparse measurement points, and extrapolation beyond the initial field. On one hand, the occurrence of transonic flows often implies the existence of special solutions, such as shock waves and other discontinuities. Shock waves, as abrupt changes in the flow field, cannot be described by differential equations. If they are forcibly included in the loss function of PINN, it will cause oscillations and discontinuities in the loss value. Therefore, the calculation of transport across shock waves becomes less reliable. On the other hand, limited by the bottleneck of non-contact measurements, the number of measurement points is scarce, and the test data is often distributed on a two-dimensional plane or a one-dimensional line. This means that PINN cannot accurately fit the three-dimensional space using these measurement points.

In this paper, we optimize PINN for such application scenarios by introducing finite difference numerical schemes to add hard constraints. This enables PINN to have both the computational characteristics of soft-constrained PINN and discrete numerical formats. The concept of discrete order [41] is effective for this type of PINN. On one hand, increasing the order means increasing the spatiotemporal coordinates involved in the operation, making the spatial range of information interaction wider [42][43][44][45], enhancing the information propagation speed in PINN, and speeding up the training process. On the other hand, increasing the order means improving the spatial wave number resolution [46][47][48], which allows for the use of fewer measurement data points for physical field inversion. Therefore, in this paper, by introducing a 7th-order WENO-Z-based finite difference discretization format [49][50][51], we generalize the technology of hard physical constraints to higher orders, achieving the modeling of physical fields relying only on a small amount of actual measurement data. Facing the strong compressibility of fluids, WENO-Z is a discontinuity detection method suitable for finite difference formats of compressible fluid shock wave problems. In the face of shock waves and similar discontinuities, WENO-Z will actively change the template configuration to avoid the non-differentiability caused by the discontinuity of the physical field as much as possible. In addition, unlike the idea of finite volume regional integration [52], the finite difference method considers that the spatial coordinate points involved in the calculation represent the local true physical values [53][54], rather than some kind of integral average. On one hand, this idea provides the possibility for the precise fusion of measurement data. On the other hand, when the resolution of the measurement points is too low, the finite difference method provides the possibility for the precise description of the fluid-solid heat transfer interface.

The following essay will be divided into three main sections for explanation: The first section introduces the method, which begins by explaining the working principle of traditional PINN and the inherent flaws from a theoretical perspective. It then discusses how Hodd-PINN addresses these flaws in the convective term, demonstrating its effectiveness and revealing hidden, deeper spurious discontinuity issues on the basis of pure convective problems. Afterwards, the solution to these spurious discontinuity problems is presented. Finally, the section describes how Hodd-PINN handles the viscous term. The second section examines the performance of Hodd-PINN, including improvements in accuracy for pure convective problems, spurious discontinuity problems, and two-dimensional heat conduction problems. The third section consists of analysis and conclusions.

# Methodology

## PINN

As a fundamental function fitting tool and interpolation method, Neural Network (NN) is the main means of data fusion in PINN. To ensure the reliability of the results in this paper, the constructed NNs all adopt the Resnet[55] technique to minimize the impact of NN characteristics on the experimental results. Each residual block consists of 2 layers of linear transformation and 2 layers of activation function, with a width of 32 for all residual blocks. In addition, considering the smoothness of the real physical field and the difficulty of training neural networks, the Tanh[56] function is uniformly used as the activation function in this paper. The neural network structure used in this paper is shown in Figure 1, where the input is the spatiotemporal coordinates of a point in the physical field, and the output is the corresponding value of the physical field at that position. For one-dimensional unsteady convection problem, the input is $(x, t)$ and the output is velocity $U$. For the two-dimensional steady-state heat conduction problem, the input is $(x, y)$ and the output is temperature $T$. Each spatiotemporal coordinate used for training is called a configuration point.

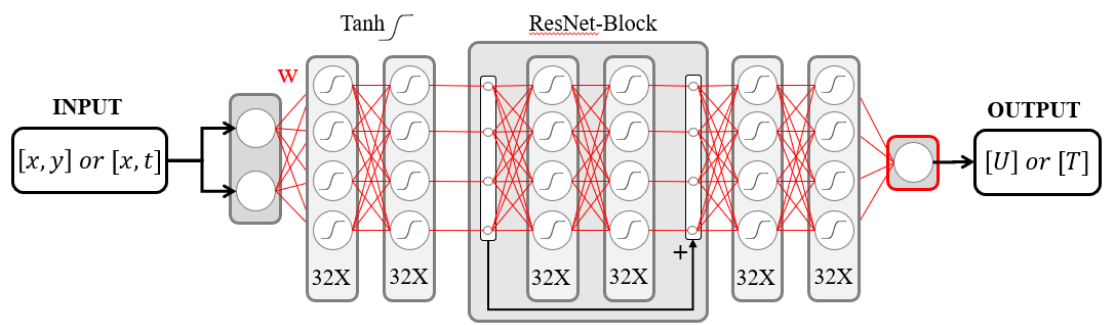

Figure 1 Illustrates the neural network structure and operation employed in this paper. In this figure, INPUT refers to the input of the neural network, while OUTPUT represents the output of the neural network. Each layer of the neural network contains a weight matrix w and bias

b, with b not explicitly labeled.

In PINN, apart from the configuration of the neural network, the training data is also a very important part. Generally speaking, there are three types of data used for training in PINN: the first type is data with specific spatiotemporal physical values known, such as temperature and velocity obtained from measurements. The second type is boundary conditions of a certain physical property with a specific spatiotemporal position, such as adiabatic boundary conditions and isothermal flow conditions. The third type is conditions of a certain physical law with a specific spatiotemporal position, such as the requirement to satisfy the Navier-Stokes equation at any point in the flow field. Among these three types of data, the first type is obtained through experimental sampling, so the quantity is limited. The second type of data depends on the nature of the boundary conditions, such as velocity boundary conditions and temperature boundary conditions obtained through experimental sampling. Adiabatic boundary conditions and isothermal flow boundary conditions can be sampled arbitrarily. The third type of data can be sampled arbitrarily, so the quantity is unlimited.

From here, it can be seen that in the PINN method, the points that can only be obtained through experimental sampling are the most precious: the values of these points can directly determine the final solution of the physical field. However, an undeniable fact is that the quantity of such points cannot be infinitely many. Especially for boundaries, if the boundary conditions at a position are not clear, then theoretically speaking, this would mean that the entire solution has a free boundary. In other words, the results obtained by PINN in this case (also in most commonly occurring cases), even if it performs the best, are just one of the numerous possible solutions.

When there are a sufficient number of sampling points, due to the neural network's inherent ability for stable generalization, the above-mentioned problem can be basically ignored. However, when there are not enough sampling points, it is necessary to make more effective use of these precious sampling points. To achieve this, we need to start solving the problem from the physical constraints part of PINN, that is, by introducing the idea of high-order finite difference. The basic idea of finite difference solving method is to find a possible solution between a few known values at specific positions in the known space, so that the partial differential equations are satisfied. The error of this solution is controlled by Taylor series expansion, providing a relatively rigorous mathematical expression. Therefore, high-order finite difference often has higher determinism and can make more effective use of the precious known information. The following section will introduce the implementation of the convection and diffusion terms based on the high-order finite difference method and complete the corresponding tests.

## The 7th-order convective term

Unlike traditional PINN, this paper does not use the direct form of partial differential equations to train the neural network but introduces the concept of high-order spatial discretization. The 7th order convection finite difference calculation method can be decomposed into three parts: 7th order positive and negative flux construction (Appendix 1), nonlinear weight construction (Appendix 2), and flux difference. By using the positive and negative flux construction (Appendix 1) and the nonlinear weight construction (Appendix 2), $\hat{u}^+_{i+\frac{1}{2}}, \hat{u}^-_{i-\frac{1}{2}}$ shown in formula 1 can be obtained, and then the Roe average[57] upwind flux difference in formula 2 can be performed.

$$\hat{u}^+_{i+\frac{1}{2}} = \sum \omega^+_m \hat{u}^+_m$$
$$\hat{u}^-_{i-\frac{1}{2}} = \sum \omega^-_m \hat{u}^-_m \quad (1)$$

$$u^{Roe}_{i+\frac{1}{2}} = \frac{\sqrt{\rho_i} u_i + \sqrt{\rho_{i+1}} u_{i+1}}{\sqrt{\rho_i} + \sqrt{\rho_{i+1}}} \quad (2)$$

To facilitate data parallelism, the flux $\hat{u}_{i+\frac{1}{2}}$ is determined by equation 3, where $sign(x)$ represents the sign function. Finally, the convective term is determined by equation 4.

$$\hat{u}_{i+\frac{1}{2}} = \frac{\left[1 + sign\left(u^{Roe}_{i+\frac{1}{2}}\right)\right]}{2} \hat{u}^+_{i+\frac{1}{2}} + \frac{\left[1 - sign\left(u^{Roe}_{i+\frac{1}{2}}\right)\right]}{2} \hat{u}^-_{i+\frac{1}{2}} \quad (3)$$

$$u \frac{\partial u}{\partial x_i} = u \frac{\left(\hat{u}_{i+\frac{1}{2}} - \hat{u}_{i-\frac{1}{2}}\right)}{\Delta x} + o(\Delta^7) \quad (4)$$

Next, taking the transient problem in one-dimensional space as an example, we will explain how to embed the above finite difference calculation method into PINN. For the time advancement process of a flow field, the PINN velocity prediction result $u^{t0}$ is used as the input for the previous time step. This input is then randomly sampled in time and space to obtain multiple configuration points with equal intervals on the spatial coordinates, thus adapting the finite difference method. The values of these configuration points are outputted after time advancement using a 7th-order finite difference format, obtaining the predicted result $u^{t1}$ for the next time step. This result is then used to train the PINN for predicting the subsequent time step. This process is shown in Figure 2. This sampling before training approach can avoid the problem of discontinuity that traditional continuous space PINN encounters: the space represented by NN is still continuous, and its advantages still exist, but the problem caused by the discontinuity of the flow field can be solved by finite difference.

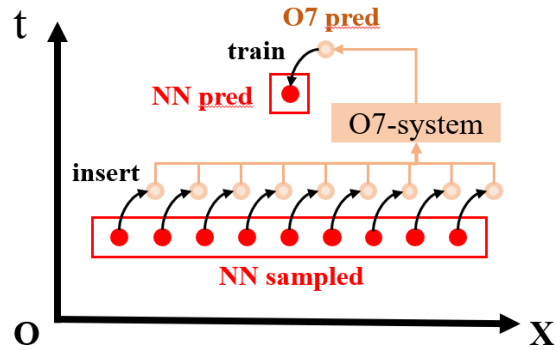

Figure 2 Illustrates the training method of Hodd-PINN using a 7th order finite difference discretization system. The red dots represent the sampling results of the neural network (NN) in the PINN. The pink area represents the structured 7th order finite difference computation system.

It should be noted that in high-order finite difference discretization processes, there exist coefficient matrices with both large and small values. If backpropagation passes through these matrices, for some special cases, such as relatively long time steps or severe oscillations in PINN's predicted results, it may lead to unstable training. Therefore, when using high-order finite difference discretization to guide PINN training, it is recommended to block this part of the backpropagation channel. In our test results, blocking did not lead to a decrease in performance, but rather improved computational speed.

Compared to traditional PINN, the significant advantage of this method is that it introduces long-distance spatial correlations between multiple configuration points, as shown in Figure 3, by utilizing the finite difference assumption. In traditional PINN, these configuration points only need to satisfy the regularity of their respective positions, but cannot guarantee the accuracy of generalization positions. However, with the introduction of numerical schemes, the values of these position points need to jointly satisfy the physical constraints, with an error only in the truncation of the Taylor expansion. This means that the introduction of numerical schemes makes the predicted physical field by PINN more controllable. Based on this, the numerical scheme introduces conservation to the predicted results of PINN through reconstruction, which is crucial in convective problems.

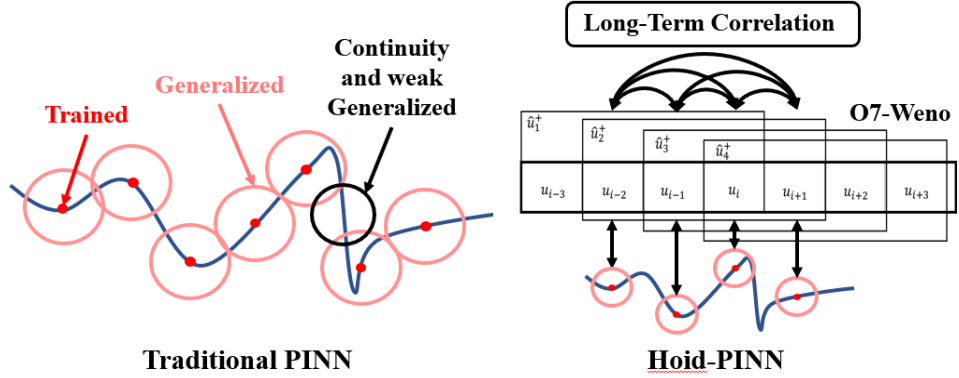

Figure 2 In the training and generalization mode of the traditional physics-informed neural network (PINN), the correlation between data points is constructed by relying on the neural network (left). Once the 7th order finite difference computation system is introduced, the correlation between data points needs to further adhere to the long-range constraints imposed by the finite difference scheme.

During the training process, first construct the PINN neural network P. Then, for each configuration point, sample the surrounding spatiotemporal coordinates at equal intervals. The sampled coordinates are then fed into network P to obtain velocity prediction results $\widehat{U}$. Next, $\widehat{U}$ is input into a 7th-order convection format for reconstruction, resulting in the physically constrained predicted $\widehat{U}^*$. Finally, by performing optimization $argmin\left(Loss(\widehat{U}, \widehat{U}^*)\right)$, the numerically conserved reconstructed results are introduced into the training process of PINN. This process is illustrated in Figure 4.

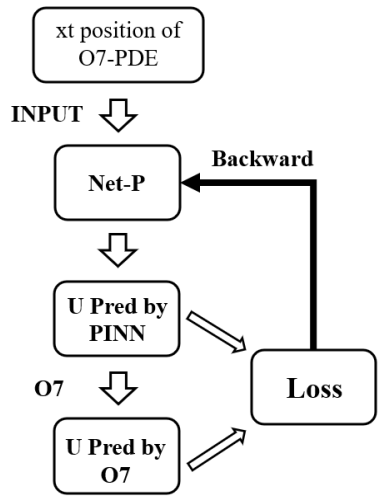

Figure 4 Training method of PINN with the introduction of O7-PDE.

To verify the proposed method in this article, a one-dimensional nonlinear convection problem is introduced, and the governing equation is determined by Equation 5.

$$\frac{\partial u}{\partial t} + \frac{\partial uu}{\partial x} = 0 \tag{5}$$

The initial setup is shown as follows in Equation 6:

$$u_{t0} = \begin{cases} 0, & x \in [0, 0.3) \\ 1, & x \in [0.3, 0.6) \\ 0, & x \in [0.6, 1] \end{cases} \tag{6}$$

The time progression range is 0.2s, and the theoretical solution to this problem is expressed in equation 7, as shown in Figure 5:

$$u_t = \begin{cases} 0, & x \in [0, 0.3) \\ \dfrac{x - 0.3}{2t}, & x \in [03, 0.3 + 2t) \\ 1, & x \in [0.3 + 2t, 0.6 + t) \\ 0, & x \in [0.6 + t, 1] \end{cases} \tag{7}$$

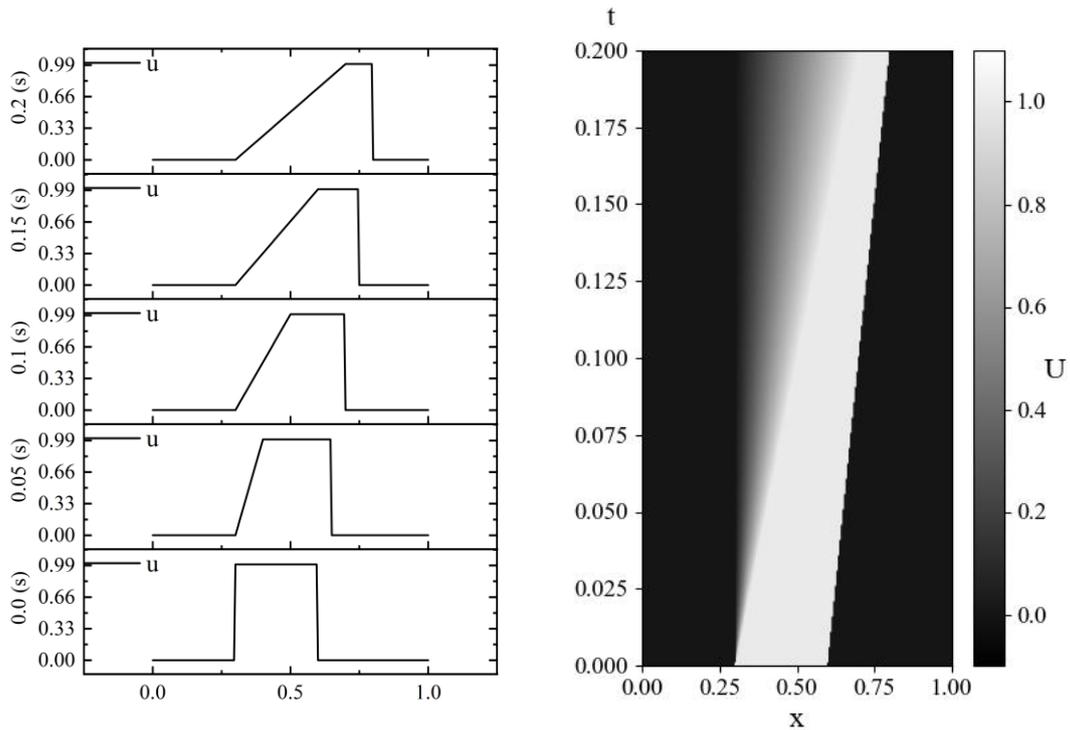

Figure 5 The theoretical solution (7) describes the pure convection problem described in equations (5) and (6).

For the typical nonlinear convection problem adopted in this article, there will always exist two weak solutions within the given time range, namely the rarefaction wave following the shock wave and the shock wave ahead. Obviously, the position of the shock wave is non-differentiable at any moment. Mathematically, this directly leads to the failure of the partial differential equation at the shock wave position. In PINN, this causes the PDE configuration points near the shock wave to generate gradients that approach infinity. Such gradients will contaminate the training results at other positions. By introducing discrete numerical formats instead of the traditional PINN differential operator, this problem can be avoided. This article compares the prediction results of the traditional PINN, PINN guided by second-order finite difference, and PINN guided by seventh-order finite difference for this problem, as shown in Figure 6, 7, and 8.

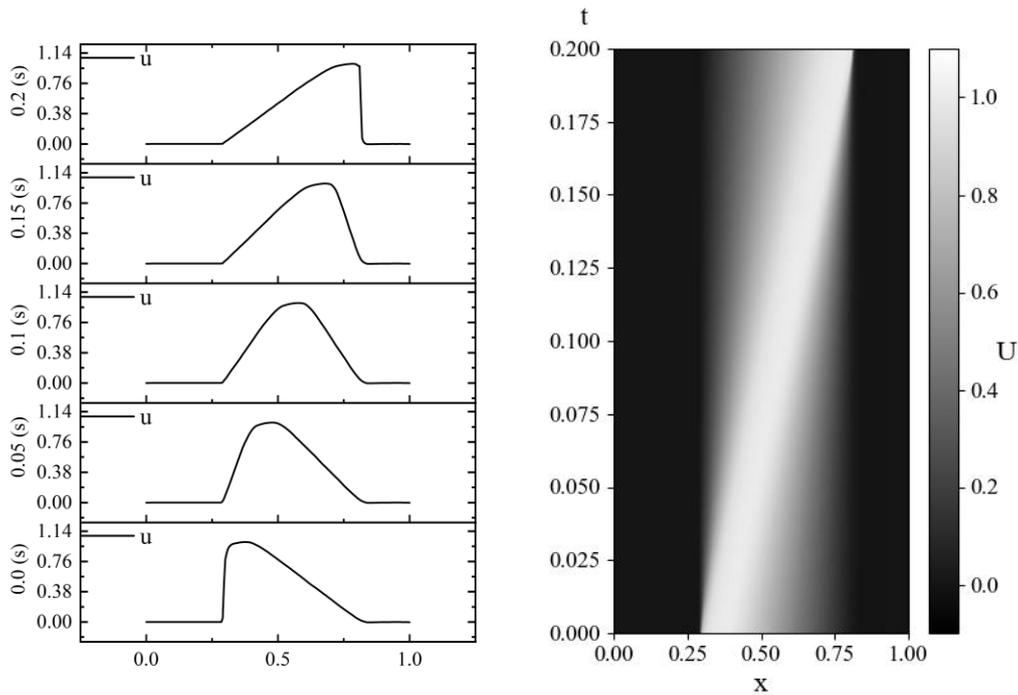

Figure 6 The predicted results after training the traditional PINN using the information provided in equations (5) and (6).

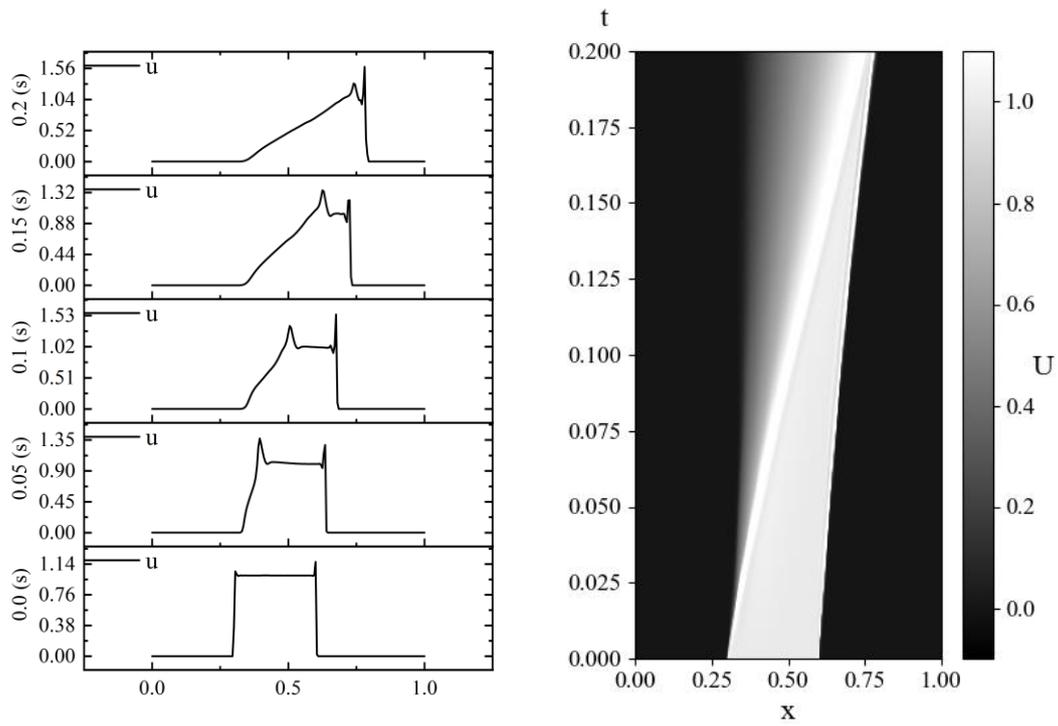

Figure 7 The prediction results of the second-order finite difference guided PINN, trained using equations (5) and (6) as known information.

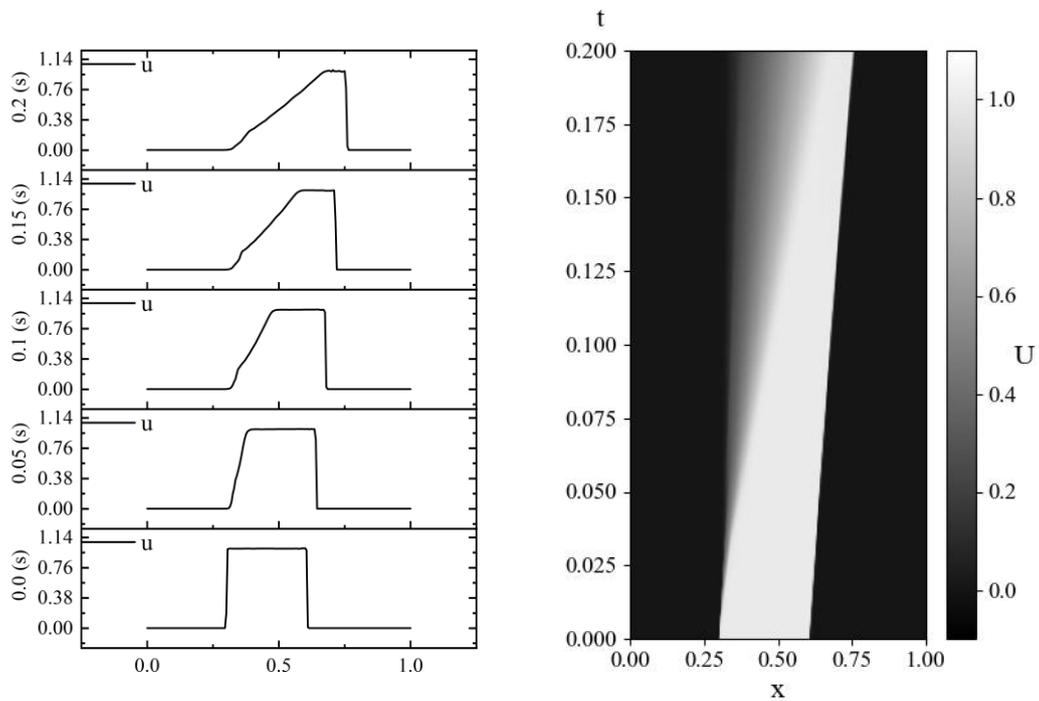

Figure 8 The predicted results of the 7th order finite difference guided PINN trained using

equations (5) and (6) as known information.

It can be observed that among the three methods, the results obtained from the 7th order finite difference guided PINN are closer to the theoretical solution. More detailed results can be found in the Result section.

However, in the case where both sparse waves and shock waves coexist, due to the fact that the penalty for shock waves is much greater than the penalty for sparse waves, there may be occasional instances where the neural network fails to detect sparse waves and predicts false discontinuities after the wave. The author believes that this problem arises because although the Mse loss function [58] can quickly reduce the overall error level, it is not very sensitive to small but important errors due to the fact that the error level spans multiple orders of magnitude. This phenomenon is similar to outliers in the training process, making it difficult to detect sparse waves[59]. This problem exists for both the second order finite difference guided PINN and the 7th order finite difference guided PINN, as shown in Figure 9.

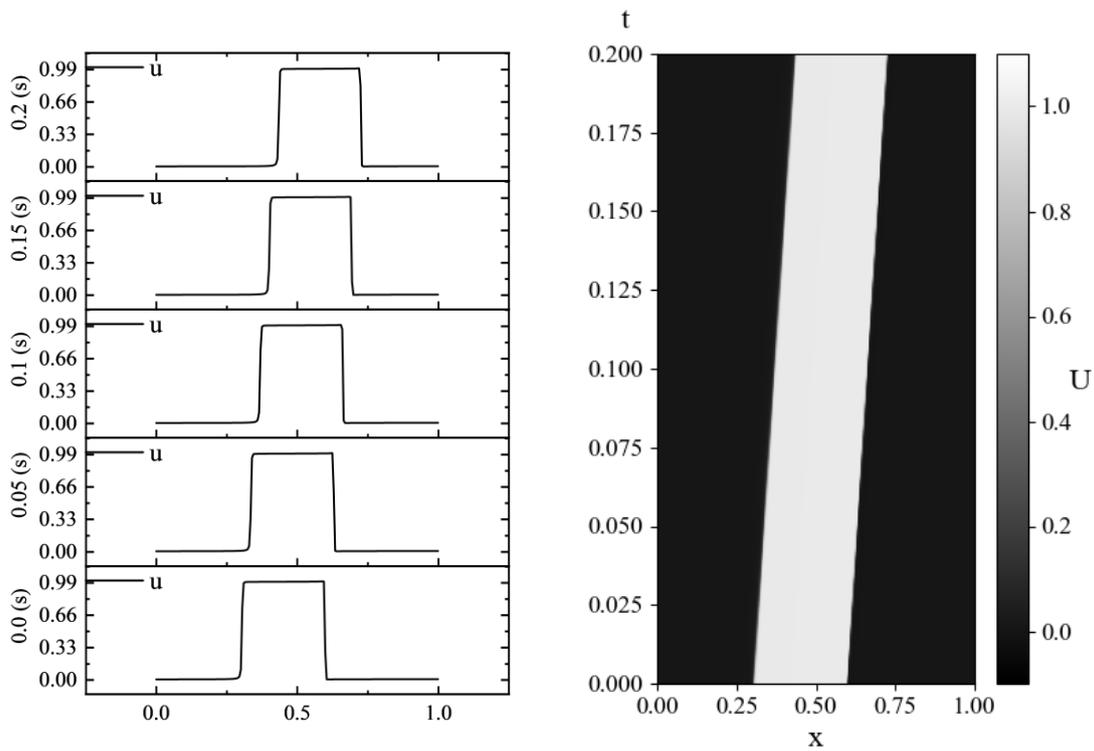

Figure 9 In some cases, the prediction results after training the PINN guided by finite difference can be observed, and it can be seen that the sparse wave on the left side was not captured by the PINN.

Generally speaking, the problem can be alleviated by adding the Mse loss function and the L1 loss function [60]. The L1 loss function can compensate for the shortcomings of Mse and better capture small errors. However, this operation does not completely solve the problem of false discontinuity in the NN fitting process, and the practice of adding the Mse loss function and the L1 loss function is not friendly to the diffusion term. In this paper, under the premise of using only the Mse loss function, the false discontinuity detection in the PINN prediction results is achieved by using the WENO discontinuity detection technique.

## WENO

First, let me introduce the implementation principle of WENO's discontinuity detection: In Appendix 2, the construction method of the smoothness factor is introduced, which realizes the discontinuity detection function of WENO. Based on this, this paper implements the pseudo-discontinuity detection in non-steady convection problems. This paper first proposes a definition of a discontinuity index:

$$\sigma = \min(softmax(IS1, IS2, IS3, IS4)) \qquad (8)$$

In general, the $\sigma$ value should be close to 0.25. The smaller the $\sigma$ value, the more likely there is a discontinuity at that position, and the stronger the discontinuity. However, in this paper, this method cannot be directly used for detecting false discontinuities in the PINN fitting results. This is because WENO is only suitable for detecting whether there is a current physical field discontinuity, and cannot be directly used to determine whether the discontinuity is reasonable. Therefore, we use traditional numerical schemes for time advancement to determine the trend of the discontinuity strength. If there is a false discontinuity at a certain position in the PINN's predicted results, then the discontinuity at that position should weaken as time advances. If a discontinuity should indeed exist at a certain position, then the strength of the discontinuity at that position is not significantly affected by time advancement.

Based on the above analysis, we define a discontinuity decay index $\beta$ based on the intermittent index $\sigma$, where $\beta = \alpha^{t+1} - \alpha^t$. Here, $\alpha^t$ is calculated from $\widehat{U}^t$, which is the result of PINN prediction, and $\alpha^{t+1}$ is the result of time advancement using the 7th-order WENO numerical scheme based on $\widehat{U}^t$. With this method, we can detect the variation trend of the initial field given by PINN in the real physical space. When a predicted discontinuity by PINN does not exist physically, the discontinuity index at that position should decrease, and $\beta$ will be significantly negative. When a predicted discontinuity by PINN actually exists, the discontinuity index at that position remains basically unchanged, and $\beta$ is basically 0. Thus, we have obtained the criteria for judging false discontinuities.

In the calculation process of this criterion, all operations are differentiable, and the constant coefficient matrices used in the calculation process are of the same order of magnitude. Therefore, we can use the WENO discontinuity detection process to achieve backpropagation. Note that it is recommended to add a layer of Relu activation function in this process. This is because in the region where the discontinuity tends to aggregate, $\beta$ may be positive. In this case, the error given by the finite difference numerical format is significant and correct, which can be captured by the loss function and does not require additional optimization through WENO discontinuity detection. This process can be expressed as shown in Figure 10.

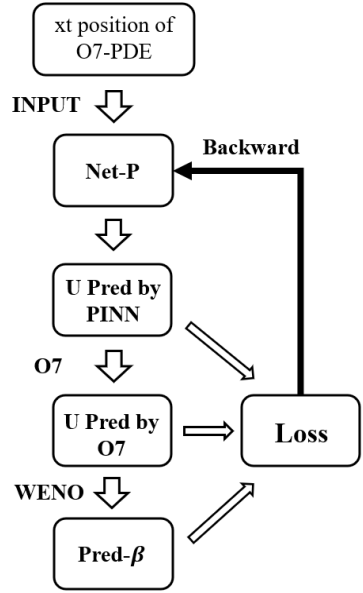

Figure 10 The training method of PINN after introducing WENO discontinuity detection.

This article compares the high-order PINN prediction results before and after adopting this technology. The comparison results can be found in the Result section.

**The 8th-order diffusion term**

In finite difference, the treatment method for the diffusion term is shown in Appendix 3. With the help of high-order discretization, we can achieve long-range propagation of limited information. In order to test the overall performance of PINN after introducing high-order discretization, this paper conducted tests on the reconstruction problem of a two-dimensional temperature field. The temperature field is a rectangular area with a length of 2pi and a width of pi, with four sides labeled as A, B, C, and D. In this temperature field, only the temperature values of a few points in the long side A are known, which belong to the Dirichlet boundary condition [61]. The remaining boundaries are known only as adiabatic boundary conditions, which belong to the Neumann boundary condition [62]. The theoretical solution for the temperature distribution on side A is:

$$T_{[0,y]} = \sin(1.5 * y) \tag{9}$$

The corresponding temperature distribution is shown in Figure 11.

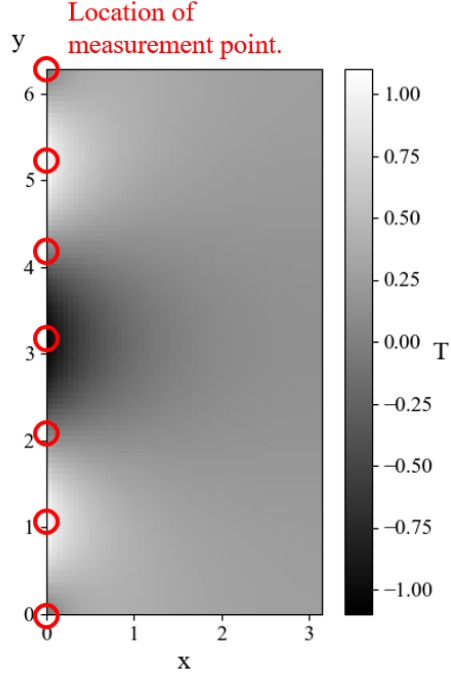

Figure 11 The target temperature field and the arrangement of temperature measurement points. Among them, the left boundary is a temperature boundary condition, while the remaining boundaries are adiabatic boundary conditions.

This article compares the traditional PINN, the PINN guided by second-order finite difference, and the PINN guided by eighth-order finite difference, respectively. The test results can be found in the Result section.

## Results

### One-dimensional convection equation

For the one-dimensional convection problem, we tested traditional PINN, PINN based on low-order discretization (O2), and PINN based on high-order discretization (O7). The tests were conducted using the Adam optimizer with a learning rate of 0.003 and the loss function consisting of MseLoss and L1Loss. The performance was recorded under different initialization seeds, different numbers of PDE configuration points, and different numbers of known value configuration points. The specific settings can be found in Table I.

Table I: Test conditions for the one-dimensional convection problem.

| PINN Method | Seed | PDE configuration points | Value configuration points |
|---|---|---|---|
| Traditional | 0 | 8000 | 50 |
| Traditional | 1 | 8000 | 50 |
| Traditional | 2 | 8000 | 50 |

| Traditional | 3 | 8000 | 50 |
| Low-order | 0 | 8000 | 50 |
| Low-order | 1 | 8000 | 50 |
| Low-order | 2 | 8000 | 50 |
| Low-order | 3 | 8000 | 50 |
| High-order | 0 | 8000 | 50 |
| High-order | 1 | 8000 | 50 |
| High-order | 2 | 8000 | 50 |
| High-order | 3 | 8000 | 50 |

In a series of tests with 8000 PDE configuration points and 50 known value configuration points, we inputted multiple seeds to test the stability of the method. The final L1 error is shown in Figure 12. Due to the significant difference in error magnitudes between different methods, we used logarithmic error to plot the boxplot in this figure.

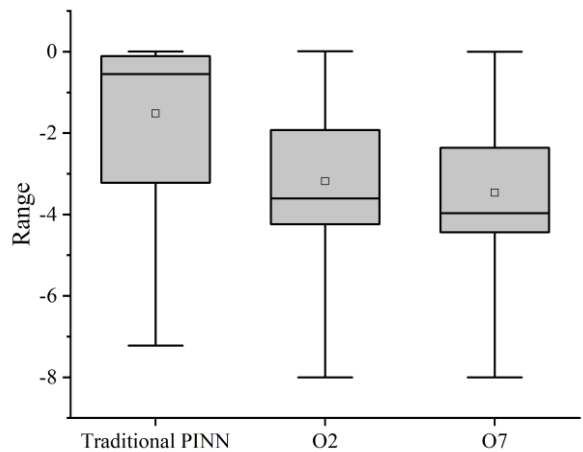

Figure 12 The error performance of traditional PINN, second-order finite difference guided PINN, and seventh-order finite difference guided PINN on the pure convection problem. The vertical axis represents the logarithm of the error.

After introducing discrete techniques into PINN, the minimum value of the absolute error decreases to the order of 10-8, which is the limit of single-precision floating-point representation. Furthermore, with the introduction of higher-order discrete techniques, the overall error level decreases. For traditional PINN, the average L1 error is approximately 0.394 m/s. After introducing O2 discrete techniques, the average L1 error decreases to 0.041 m/s. With the further introduction of higher-order discrete techniques, the average L1 error decreases to 0.037 m/s, which is an additional 9.7% reduction in error. Compared to traditional PINN, the error is reduced by a total of 90.6%.

## False positive detection

By introducing the spurious discontinuity detection technique, this study successfully improves the ability of neural networks (NN) to capture sparse waves during the training process. Tests were conducted to compare the performance of PINN before and after the introduction of the spurious discontinuity detection technique. These tests used the Adam optimizer with a learning rate of 0.003 and the MseLoss as the loss function, while keeping the other parameters the same. The decrease in L1 error between the predicted solution and the theoretical solution with respect to the number of training steps is shown in Figure 13.

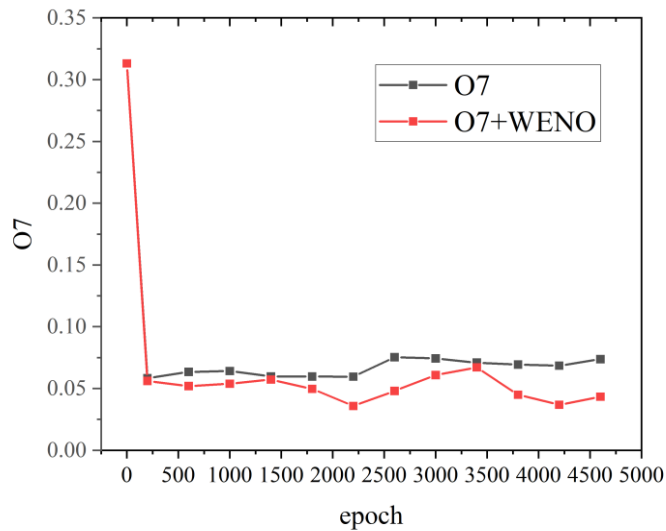

Figure 13 The training process error performance of the 7th order finite difference guided PINN before and after the introduction of WENO pseudo-discontinuity detection technique.

It can be clearly observed that after the introduction of the pseudo-discontinuity detection technique, the error between the predicted results of the PINN and the theoretical solution is smaller. Excluding the initial field error that is not trained, the introduction of the pseudo-discontinuity detection technique reduces the average error level from 0.0669 to 0.0507, resulting in a 24.2% decrease in error.

## Two-dimensional heat conduction equation

For the specified two-dimensional heat conduction problem in this article, traditional PINN, PINN guided by second-order finite difference, and PINN guided by eighth-order finite difference were tested. The tests were conducted using the Adam optimizer with a learning rate of 0.001 and the MseLoss as the loss function. The performance of the two methods was recorded under different initialization seeds, different numbers of PDE configuration points, and different numbers of known value configuration points. The specific settings can be found in Table II.

Table II: Test condition settings for the two-dimensional heat conduction problem

| PINN Method | Seed | PDE configuration points | Value configuration points | Adiabatic configuration points |
|---|---|---|---|---|
| Traditional | 0 | 800 | 7 | 200 |
| Traditional | 1 | 800 | 7 | 200 |
| Traditional | 2 | 800 | 7 | 200 |
| Traditional | 0 | 800 | 9 | 200 |
| Traditional | 0 | 800 | 15 | 200 |
| Traditional | 0 | 800 | 31 | 200 |
| Traditional | 0 | 8000 | 400 | 400 |
| Low-order | 0 | 800 | 7 | 200 |
| Low-order | 1 | 800 | 7 | 200 |
| Low-order | 2 | 800 | 7 | 200 |
| Low-order | 0 | 800 | 9 | 200 |
| Low-order | 0 | 800 | 15 | 200 |
| Low-order | 0 | 800 | 31 | 200 |
| High-order | 0 | 800 | 7 | 200 |
| High-order | 1 | 800 | 7 | 200 |
| High-order | 2 | 800 | 7 | 200 |
| High-order | 0 | 800 | 9 | 200 |
| High-order | 0 | 800 | 15 | 200 |
| High-order | 0 | 800 | 31 | 200 |

In the case of only 7 known value configuration points, we conducted comprehensive testing using multiple seeds, and the logarithmic error is shown in Figure 14:

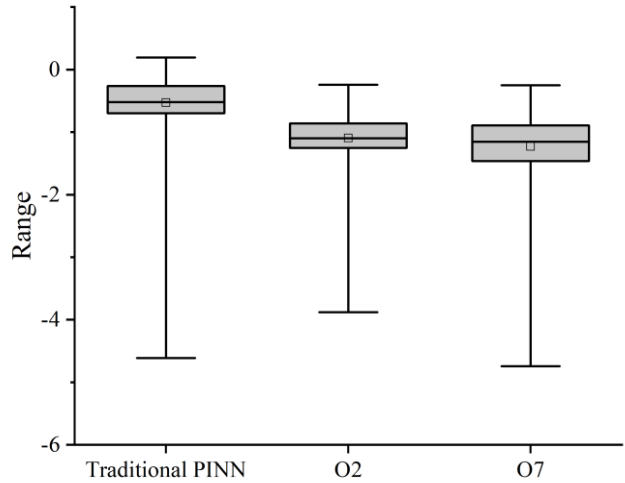

Figure 14: Error performance of traditional PINN, second-order finite difference guided PINN, and high-order finite difference guided PINN on the pure heat conduction problem, with the vertical axis representing the logarithm of the error.

From Figure 14, it can be observed that the 8th-order finite difference guided PINN has a significantly lower error level compared to the traditional PINN and the second-order finite difference guided PINN. The average error of the traditional PINN is approximately 0.434, the average error of the second-order finite difference guided PINN is approximately 0.101, and the average error of the 8th-order finite difference guided PINN is approximately 0.088. Compared to the second-order, it has improved by about 12.8%, and compared to the traditional PINN, it has improved by about 34.6%.

## Conclusion

This article combines traditional finite difference numerical schemes, WENO discontinuity detection, and traditional PINN methods to develop a high-order discontinuity detection PINN that can achieve inversion of velocity and temperature fields with a small number of measurement points. By introducing high-order finite difference methods, the efficiency of utilizing known data points by PINN is effectively improved. By introducing a pseudo-discontinuity detection method, the ability of PINN to capture sparse waves is enhanced. Through this study, the following typical conclusions can be obtained:

1. Traditional PINN cannot solve discontinuities, which is caused by the inherent continuous function representation capability of NN. Finite difference methods can replace the automatic differentiation process of NN, bypassing the problems caused by discontinuities while retaining the advantages of NN's continuity.

2. Introducing finite difference methods can effectively solve the problem of

insufficient known data points, and under the same amount of data, higher-order schemes perform better than lower-order schemes. For the convection term, the 7th order scheme improves by an average of 9.7% compared to the 2nd order scheme, and for the diffusion term, the 7th order scheme improves by an average of 12.8% compared to the 2nd order scheme.

3.The pseudo-discontinuity detection technique based on WENO developed in this article can effectively identify pseudo-discontinuities in the fitting results of NN and eliminate them through optimization. This leads to a 24.2% improvement in accuracy.

## Author's contribution

Ruquan You: Supervision, Conceptualization, Data curation, Writing-review & editing. Shuming Zhang: Formal analysis, Investigation, Writing-original draft. Tinglin Kong: Provide partial formula derivation. Haiwang Li: Supervision, Conceptualization.

## Acknowlegments

This work is supported by Beijing Nova Program; The Advanced Aviation Power Innovation Workstation (established by the China Aviation Engine Research Institute) [grant number HKCX2022-01-007]; The Aeronautical Science Fund (provided by Aviation Industry Corporation of China)[grant number 2022Z039051006];The funding sources are not involved in study design; in the collection, analysis and interpretation of data; in the writing of the report; and in the decision to submit the article for publication.

# Appendix 1: Construction method for positive and negative flux of 7th order.

The difference of the convective term needs to be completed through flux reconstruction. Taking the construction of the derivative $\frac{\partial u}{\partial x}$ of the velocity component $u$ as an example, the flux reconstruction method is introduced below. The derivative $\frac{\partial u}{\partial x_i}$ at grid point $i(i = 0,1 \ldots N)$ is written in the form of a $k$th-order accuracy flux, that is:

$$\frac{\partial u}{\partial x_i} = \frac{\hat{u}_{i+\frac{1}{2}} - \hat{u}_{i-\frac{1}{2}}}{\Delta x} + o(\Delta x^k) \qquad (10)$$

Given the flux $\hat{u}$ distributed on $x$ as $\hat{u}(x)$, define the function $F(x)$ on $x$.

$$F(x) = \frac{1}{\Delta x} \int_{x-\frac{1}{2}\Delta x}^{x+\frac{1}{2}\Delta x} \hat{u}(r) dr \qquad (11)$$

$$F(x) = u(x) + o(\Delta x^{k+1}) \qquad (12)$$

Substituting equation 12 into equation 11 and differentiating both sides:

$$\frac{\partial u}{\partial x} + o(\Delta x^k) = \frac{\hat{u}\left(x + \frac{1}{2}\Delta x\right) - \hat{u}\left(x - \frac{1}{2}\Delta x\right)}{\Delta x} \qquad (13)$$

Equation 13 is equal to Equation 10 at $x = x_i$. For the upwind flux reconstruction, positive and negative fluxes need to be constructed at each grid boundary, using different templates. Figure 15 shows the templates for constructing positive and negative fluxes at $x_{i+\frac{1}{2}}$ using the seventh-order upwind WENO format.

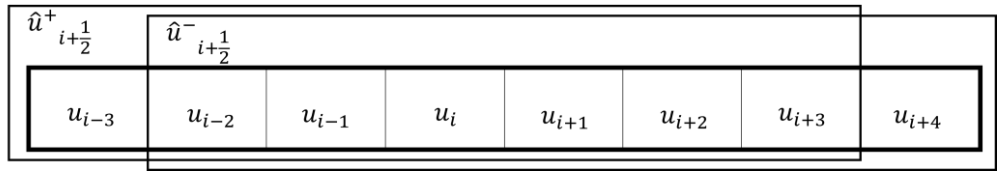

Figure 15　Seven-order windward format positive and negative flux template.

Next, taking the construction of the positive flux $\hat{u}^+_{i+\frac{1}{2}}$ at the position $x_{i+\frac{1}{2}}$ as an example, the seventh-order WENO reconstruction is introduced. The seventh-order WENO reconstruction uses a total of 7 points for a positive/negative flux template, which is split into 4 fourth-order templates and then weighted averaged, as shown in Figure 16.

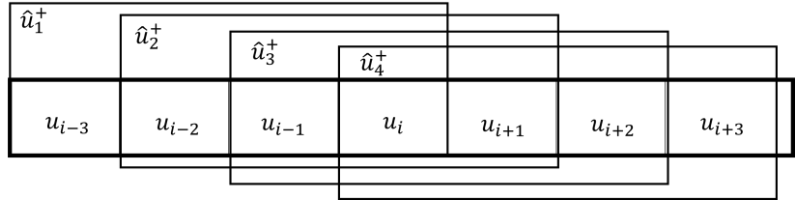

Figure 16: Composition of the fourth-order template for the WENO scheme.

Where:

$$\hat{u}^+_{i+\frac{1}{2}} = \omega_1^+ \hat{u}_1^+ + \omega_2^+ \hat{u}_2^+ + \omega_3^+ \hat{u}_3^+ + \omega_4^+ \hat{u}_4^+ \tag{14}$$

Next, we will introduce how to construct these four fourth-order templates. Let $\hat{u}_1(x), \hat{u}_2(x), \hat{u}_3(x), \hat{u}_4(x)$ be four cubic polynomials, namely:

$$\hat{u}_m(x) = a_m + b_m x + c_m x^2 + d_m x^3, (m = 1,2,3,4) \tag{15}$$

Follows:

$$\frac{1}{\Delta x} \int_{x_{i-5+m+r}-\frac{\Delta x}{2}}^{x_{i-5+m+r}+\frac{\Delta x}{2}} \hat{u}_m(x)dx = u_{i-5+m+r}, (r = 1,2,3,4) \tag{16}$$

Which can be written as:

$$\int_{-\frac{11}{2}+m+r}^{-\frac{9}{2}+m+r} \hat{u}_m(\beta)dx = u_{i-5+m+r}, (r = 1,2,3,4) \tag{17}$$

Where:

$$\beta = \frac{x - x_i}{\Delta x} \tag{18}$$

From this, we can obtain a set of linear equations and solve them, thus using $u_j, (j = i - 3, i - 2, \ldots, i + 3)$ to represent $a_m, b_m, c_m, d_m (m = 1,2,3,4)$.

Substituting $\beta = \frac{1}{2}$ into $\hat{u}_m(\beta), (m = 1,2,3,4)$ yields:

$$\hat{u}_1^+ = -\frac{1}{4}u_{i-3} + \frac{13}{12}u_{i-2} - \frac{23}{12}u_{i-1} + \frac{25}{12}u_i$$
$$\hat{u}_2^+ = \frac{1}{12}u_{i-2} - \frac{5}{12}u_{i-1} + \frac{13}{12}u_i + \frac{1}{4}u_{i+1}$$
$$\hat{u}_3^+ = -\frac{1}{12}u_{i-1} + \frac{7}{12}u_i + \frac{7}{12}u_{i+1} - \frac{1}{12}u_{i+2}$$
$$\hat{u}_4^+ = \frac{1}{4}u_{i-1} + \frac{13}{12}u_i - \frac{5}{12}u_{i+1} + \frac{1}{12}u_{i+2}$$
(19)

Substituting $\beta = -\frac{1}{2}$ into $\hat{u}_m(\beta), (m = 1,2,3,4)$ yields:

$$\hat{u}_1^- = \frac{1}{12}u_{i-3} - \frac{5}{12}u_{i-2} + \frac{13}{12}u_{i-1} + \frac{1}{4}u_i$$
$$\hat{u}_2^- = -\frac{1}{12}u_{i-2} + \frac{7}{12}u_{i-1} + \frac{7}{12}u_i - \frac{1}{12}u_{i+1}$$
$$\hat{u}_3^- = \frac{1}{4}u_{i-1} + \frac{13}{12}u_i - \frac{5}{12}u_{i+1} + \frac{1}{12}u_{i+2}$$
$$\hat{u}_4^- = \frac{25}{12}u_{i-1} - \frac{23}{12}u_i + \frac{13}{12}u_{i+1} - \frac{1}{4}u_{i+2}$$
(20)

At this point, the construction of the four fourth-order templates for the seventh-order scheme is completed. The positive and negative fluxes $\hat{u}^+_{i+\frac{1}{2}}, \hat{u}^-_{i-\frac{1}{2}}$ for the seventh-order scheme will be obtained by averaging the four fourth-order positive and negative fluxes. This part will be explained in Appendix 2.

## Appendix 2: Nonlinear Weight Construction Method

This section will introduce the determination method of the nonlinear weights $\omega_m^\pm, (m = 1,2,3,4)$. Constructing nonlinear weights first requires determining the linear seventh-order template, then determining the linear weights $\gamma_m^\pm, (m = 1,2,3,4)$, and finally determining the nonlinear weights $\omega_m^\pm, (m = 1,2,3,4)$. Similarly to the fourth-order template, let $\hat{u}(x)$ be a sixth-degree polynomial:

$$\hat{u}(x) = a + b\,x + c\,x^2 + d\,x^3 + e\,x^4 + f\,x^5 + g\,x^6, (m = 1,2,3,4) \quad (21)$$

This equation satisfies:

$$\int_{-\frac{9}{2}+r}^{-\frac{7}{2}+r} \hat{u}(\beta)d\beta = u_{i-4+r}, (r = 1,2,3,4,5,6,7) \quad (22)$$

Where $\beta$ is determined by equation 18, thus using $u_j, (j = i - 3, i - 2, \ldots, i +$

3) to represent $a, b, c, d, e, f, g$. Substituting $\beta = \frac{1}{2}$ into $\hat{u}(\beta)$ yields:

$$\hat{u}^+_{i+\frac{1}{2}} = -\frac{1}{140}u_{i-3} + \frac{5}{84}u_{i-2} - \frac{101}{420}u_{i-1} + \frac{319}{420}u_i + \frac{107}{210}u_{i+1} - \frac{19}{210}u_{i+2} + \frac{1}{105}u_{i+3} \quad (23)$$

And because:

$$\hat{u}^+_{i+\frac{1}{2}} = \hat{u}^+_m \gamma^+_m, (m = 1,2,3,4) \quad (24)$$

Substituting equations 19 and 23 into equation 24, we can solve for the linear weights:

$$(\gamma^+_1, \gamma^+_2, \gamma^+_3, \gamma^+_4) = \left(\frac{1}{35}, \frac{12}{35}, \frac{18}{35}, \frac{4}{35}\right) \quad (25)$$

Substituting $\beta = -\frac{1}{2}$ into $\hat{u}(\beta)$, we obtain:

$$(\gamma^-_1, \gamma^-_2, \gamma^-_3, \gamma^-_4) = \left(\frac{4}{35}, \frac{18}{35}, \frac{12}{35}, \frac{1}{35}\right) \quad (26)$$

The nonlinear weights are constructed from the linear weights and a smoothing factor. The smoothing factor should satisfy the condition that it is small in regions where $\hat{u}_m(x)$ is completely smooth, and large in regions with discontinuities. In this paper, the smoothing factor $IS_m$ is implemented using the scheme proposed by C.-W. Shu and S. Osher:

$$IS_m = \sum_{l=1}^{3} \int_{x_i-\frac{1}{2}\Delta x}^{x_i+\frac{1}{2}\Delta x} \Delta x^{2l-1} \left(\frac{d^l \hat{u}_m(x)}{dx^l}\right)^2 dx \quad (27)$$

By substituting equations 20 to 23 into equation 15, and then substituting the result into equation 27, we can solve for the smoothing factors $IS_1, IS_2, IS_3, IS_4$. According to the WENO-Z scheme proposed by R. Borges et al., we need to construct a factor $\sigma$ that measures the second-order moment of the entire template. The $u_j, j = (i-3, i-2, \ldots i+3)$ used in $IS_m$ can be Taylor expanded up to seventh order at $x = x_i$, while $\sigma$ should be a linear combination of $IS_m$ and of the order $o(\Delta x^7)$, yielding:

$$\sigma = IS_1 + 3IS_2 - 3IS_3 + IS_4$$
$$= \left(-\frac{2}{3}\frac{\partial u}{\partial x}\frac{\partial^6 u}{\partial x^6} + \frac{13}{3}\frac{\partial^2 u}{\partial x^2}\frac{\partial^5 u}{\partial x^5} - \frac{1607}{120}\frac{\partial^3 u}{\partial x^3}\frac{\partial^4 u}{\partial x^4}\right)\Delta x^7 + o(\Delta x^8) \quad (28)$$

Then we construct the nonlinear weights $\alpha^\pm_m$:

$$\alpha^\pm_m = \gamma^+_m \left(1 + \left|\frac{\sigma}{IS_m + \varepsilon}\right|^p\right) \quad (29)$$

In the equation, ε is a small value to prevent division by zero, typically set to

$10^{-16}$, and p is an exponent that controls the strength of the ENO behavior, usually set to 1 or 2.

Finally, the $\alpha_m^{\pm}$ weights are normalized to obtain the final nonlinear weights:

$$\omega_m^{\pm} = \frac{\alpha_m^{\pm}}{\alpha_1^{\pm} + \alpha_2^{\pm} + \alpha_3^{\pm} + \alpha_4^{\pm}} \tag{30}$$

## Appendix 3, Construction Method of 8th Order Diffusion Term

In this paper, the eighth-order central differencing scheme is used to discretize the diffusion term. The goal is to obtain an eighth-order accurate approximation of the second derivative. Taking $\frac{\partial^2 u}{\partial x^2}$ as an example for the same direction second derivative, we have:

$$\frac{1}{\Delta x^2} \sum a_i u_i = \frac{\partial^2 u}{\partial x^2} + o(\Delta x^8) \tag{31}$$

To obtain the value of $a_j$, this paper uses the method of Taylor expanding $u_j$ at $x = x_i$ to the 9th order, setting the coefficient of the second-order term to 1 and the rest to 0. There are a total of 9 equations, requiring 9 degrees of freedom to determine the solution, in the format of a 9-point scheme $(j = i - 4, i - 3, ..., i + 4)$. After forming a linear system of equations, solving it will yield the expression for the 8th order diffusion term.

$$\frac{\partial^2 u}{\partial x^2} = \frac{-\frac{1}{560}u_{i-4} + \frac{8}{315}u_{i-3} - \frac{1}{5}u_{i-2} + \frac{8}{5}u_{i-1} - \frac{205}{72}u_i + \frac{8}{5}u_{i+1} - \frac{1}{5}u_{i+2} + \frac{8}{315}u_{i+3} - \frac{1}{560}u_{i+4}}{\Delta x^2}$$

$$+ o(\Delta x^8) \tag{32}$$